\documentclass[aps,prb,twocolumn,groupedaddress,showpacs]{revtex4}
\usepackage{graphicx,amssymb,amsmath}

\newcommand{\rf}[1]{(\ref{#1})}
\newcommand{\al}[1]{\begin{aligned}#1\end{aligned}}
\newcommand{\eq}[1]{\begin{equation}#1\end{equation}}

\begin{document}

\title{Electron Transport Driven by Nonequilibrium Magnetic Textures}

\author{Yaroslav Tserkovnyak}
\author{Matthew Mecklenburg}
\affiliation{Department of Physics and Astronomy, University of California, Los Angeles, California 90095, USA}

\date{\today}

\begin{abstract}
Spin-polarized electron transport driven by inhomogeneous magnetic dynamics is discussed in the limit of a large exchange coupling. Electron spins rigidly following the time-dependent magnetic profile experience spin-dependent fictitious electric and magnetic fields. We show that the electric field acquires important corrections due to spin dephasing, when one relaxes the spin-projection approximation. Furthermore, spin-flip scattering between the spin bands needs to be taken into account in order to calculate voltages and spin accumulations induced by the magnetic dynamics. A phenomenological approach based on the Onsager reciprocity principle is developed, which allows us to capture the effect of spin dephasing and make a connection to the well studied problem of current-driven magnetic dynamics. A number of results that recently appeared in 
the literature are related and generalized.
\end{abstract}

\pacs{72.15.Gd,72.25.Ba,75.47.-m,75.75.+a}


\maketitle

Interest in magnetic heterostructures,\cite{ralphJMMM07} which was initially fueled by the discovery of the giant magnetoresistance and, a decade later, by the current-induced switching in spin valves and related systems, has more recently spilled over into current-driven phenomena in magnetic bulk, individual magnetic films, and nanowires.\cite{tserkovJMMM08} Particular attention was given to the problems of current-driven Doppler shift of spin waves, magnetic instabilities, and domain-wall motion. The latter has also enjoyed a very vibrant experimental activity, which is in part motivated by a promising application potential in spintronics. The past year\cite{barnesPRL07,duinePRB08,saslowPRB07,yangCM07} saw a revival of interest in the inverse effect of electromotive forces induced by the time-dependent magnetization, which were previously studied in various physical contexts (see, e.g., Refs.~\onlinecite{volovikJPC87,sternPRL92,stonePRB96}). In this paper, we will exploit the reciprocal relation between the two phenomena, which will allow us to understand important spin-dephasing corrections to the electromotive force. Such corrections were first mentioned in Ref.~\onlinecite{duinePRB08} and the Onsager principle in the present context was invoked in Ref.~\onlinecite{saslowPRB07}. Reference~\onlinecite{barnesPRL07} reported the magnetically-induced electromotive forces as a manifestation of the position-dependent Berry phase accumulation and Ref.~\onlinecite{yangCM07} considered these forces acting on semiclassical wave packet motion, mainly reproducing results from Ref.~\onlinecite{volovikJPC87}. For completeness, it should also be mentioned that a much earlier paper\cite{bergerPRB86} already contains some seminal phenomenological insights related to the problem of the electric response to the magnetic domain-wall dynamics.

In the following, we start by recalling how most of the results recently discussed in the literature can be captured by an SU(2) gauge transformation together with the projection of spins on the magnetic direction.\cite{volovikJPC87} The corrections due to the remaining transverse spin dynamics are governed by spin dephasing, which have already been studied for the reciprocal process of current-driven magnetic dynamics,\cite{tserkovJMMM08} and can be translated to the current problem by the Onsager principle. We will develop a general framework, that will allow us to relate and generalize the more specialized cases, which were recently studied using different methods.\cite{barnesPRL07,duinePRB08,saslowPRB07,yangCM07} Finally, we will derive spin-charge diffusion equations, accounting for spin-flip scattering and respecting local charge neutrality, which is necessary in order to relate the microscopic electromotive forces to measurable quantities, such as an induced voltage and spin accumulation.

Most of our analysis will pertain to the following time-dependent Hamiltonian:
\eq{H(t)=\frac{p^2}{2m}+\frac{\Delta_{\rm xc}}{2}\boldsymbol{\hat{\sigma}}\cdot\mathbf{m}(\mathbf{r},t)+V_c(\mathbf{r},t)+H_{\sigma}\,.\label{H}}
Here, $H_\sigma$ is the contribution due to spin-relaxation processes, which will be characterized by a Bloch-type $T_1$ spin flipping and $T_2$ spin dephasing, and $V_c(\mathbf{r},t)$ stands for a Hartree charging potential, which will be taken into account only insofar as enforcing local charge neutrality. $\Delta_{\rm xc}$ is the ferromagnetic exchange band splitting, $\boldsymbol{\hat{\sigma}}$ is the vector of Pauli spin matrices, and $\mathbf{m}$ stands for the local magnetization direction unit vector, so that the magnetization is given by $\mathbf{M}=M\mathbf{m}$. The exchange field $\Delta_{\rm xc}\mathbf{m}$ may in practice be provided by localized magnetic $d$ orbitals (as in the so-called $s-d$ model) or it may self-consistently be governed by the itinerant electron spin density (as in the Stoner model or local spin-density approximation).\cite{tserkovJMMM08} We will first perform a microscopic calculation for the idealized Hamiltonian \rf{H} neglecting $H_\sigma$ and subsequently utilize the Onsager theorem to capture the spin-dephasing corrections. The spin-flip scattering will be included phenomenologically in the final diffusion equation.

By disregarding $H_\sigma$, we can perform an SU(2) gauge transformation by rotating $\mathbf{m}$ to point along the $\mathbf{z}$ axis for all $\mathbf{r}$ and $t$.\cite{volovikJPC87,tataraJPSJ07} This is conveniently achieved by the Hermitian spin-rotation matrix $\hat{U}=\boldsymbol{\hat{\sigma}}\cdot\mathbf{n}$ (such that $\hat{U}=\hat{U}^\dagger=\hat{U}^{-1}$), where $\mathbf{n}$ is the unit vector $\mathbf{n}\propto\mathbf{m}+\mathbf{z}$.  It is easy to see that $\hat{U}(\boldsymbol{\hat{\sigma}}\cdot\mathbf{m})\hat{U}=\hat{\sigma}_z$ (since $\hat{U}$ corresponds to a $\pi$-angle spin rotation around $\mathbf{n}$). By applying this gauge transformation to the spinor wave function, we get for the transformed Hamiltonian
\eq{H^\prime(t)=\frac{1}{2m}\left(\mathbf{p}-\mathbf{\hat{A}}\right)^2+\hat{V}+\frac{\Delta_{\rm xc}}{2}\hat{\sigma}_z+V_c\,,}
where the SU(2) vector potential is given by $\hat{A}_i=i\hbar\hat{U}\nabla_i\hat{U}=-\hbar\boldsymbol{\hat{\sigma}}\cdot(\mathbf{n}\times\nabla_i\mathbf{n})$ and the SU(2) ordinary potential is $\hat{V}=-i\hbar\hat{U}\partial_t\hat{U}=\hbar\boldsymbol{\hat{\sigma}}\cdot(\mathbf{n}\times\partial_t\mathbf{n})$ (setting the particle charge and speed of light to unity). $\mathbf{p}=-i\hbar\boldsymbol{\nabla}$ is the canonical momentum. If the exchange field $\Delta_{\rm xc}$ is large and the magnetic texture is sufficiently smooth and slow, we can project the fictitious potentials on the $z$ axis as $\hat{V}\to V\hat{\sigma}_z$, where $V=\hbar\mathbf{z}\cdot(\mathbf{n}\times\partial_t\mathbf{n})=\hbar\sin^2(\theta/2)\partial_t\phi$, and similarly for the vector potential, $A_i=-\hbar\mathbf{z}\cdot(\mathbf{n}\times\nabla_i\mathbf{n})=-\hbar\sin^2(\theta/2)\nabla_i\phi$, where $(\theta,\phi)$ are the spherical angles parametrizing $\mathbf{m}$. We thus get for the effective electric field\cite{volovikJPC87,duinePRB08,yangCM07}
\eq{\mathbf{E}=-\partial_t\mathbf{A}-\boldsymbol{\nabla}V=\frac{\hbar}{2}\sin\theta\left[(\partial_t\theta)(\boldsymbol{\nabla}\phi)-(\partial_t\phi)(\boldsymbol{\nabla}\theta)\right]\,,\label{E}}
or, written in the explicitly spin-rotationally-invariant form,
\eq{E_i=\frac{\hbar}{2}\mathbf{m}\cdot\left(\partial_t\mathbf{m}\times\nabla_i\mathbf{m}\right)\,.\label{Ei}}
The effective magnetic field is\cite{volovikJPC87,yePRL99,yangCM07}
\eq{\al{\mathbf{B}=\boldsymbol{\nabla}\times\mathbf{A}&=\frac{\hbar}{2}\sin\theta(\boldsymbol{\nabla}\phi)\times(\boldsymbol{\nabla}\theta)\\&=\frac{\hbar}{4}\epsilon^{ijk}m_i(\boldsymbol{\nabla}m_k\times\boldsymbol{\nabla}m_j)\,,}\label{B}}
where $\epsilon^{ijk}$ is the antisymmetric Levi-Civita tensor, and a summation over repeated indices is implied. The total force on a spin-$\uparrow$ ($\downarrow$) electron moving with velocity $\mathbf{v}$ is thus given by
\eq{\mathbf{F}^{\uparrow(\downarrow)}=\pm\left(\mathbf{E}+\mathbf{v}\times\mathbf{B}\right)\,.\label{F}}

Eqs.~\rf{E} and \rf{B} were first derived in Ref.~\onlinecite{volovikJPC87} and recently rederived within a semiclassical wave-packet analysis.\cite{yangCM07} The gauge-transformation based approach\cite{volovikJPC87} puts the result into a broader perspective, allowing us, for example, to consider the effect of the magnetic field \rf{B} on the quantum transport corrections, such as a weak localization, as well as to include spin-independent electron-electron interactions, which would not modify fictitious fields \rf{E} and \rf{B}. Note, in particular, that the magnetic field \rf{B} can in practice be quite large: For example, for a static magnetic variation on the scale of 10~nm, the corresponding fictitious field is of the order of 10~T. It is most convenient to estimate the strength of the electric field \rf{E} by the characteristic voltage it induces over a region where the magnetization direction flips its direction: $\hbar\omega/e$, where $\omega$ is the frequency of the magnetic dynamics. In the following, we will concentrate on the semiclassical spin and charge diffusion generated by the effective electric field $\mathbf{E}$. In order to make a closer connection to the experimentally relevant quantities, we will need to take into account spin relaxation and also enforce local charge neutrality for electron diffusion.

The role of spin relaxation can be twofold. First of all, spin accumulation, which will generally be generated by the spin-dependent force \rf{F} will relax, characterized by the longitudinal spin-flip time $T_1$. There is also another more subtle effect, which is due to the dephasing of electron spins following a dynamic magnetic profile, since the exchange field $\Delta_{\rm xc}$ is not infinite and spins do not perfectly align with the local magnetization. Hence, there is generally a finite spin misalignment, which dephases with a characteristic time $T_2$. This gives corrections to the results obtained by a rigid projection of spins on the local magnetization direction. We will see that such corrections turn out to be important for the currents generated by magnetic dynamics, in the same sense that analogous corrections are crucial for understanding current-induced magnetic motion.\cite{tserkovJMMM08}

Let us now take a step aside, by recalling the general expression for the dynamics of an isotropic ferromagnet well below the Curie temperature:\cite{tserkovJMMM08}
\eq{\al{\partial_t\mathbf{M}=&-\gamma\mathbf{M}\times\mathbf{H}_{\rm eff}+\frac{\alpha}{M}\mathbf{M}\times\partial_t\mathbf{M}\\&+\frac{\hbar\left(\sigma_\uparrow-\sigma_\downarrow\right)}{2S}\left(\nabla_i\mu\right)\left(1-\frac{\beta}{M}\mathbf{M}\times\right)\nabla_i\mathbf{M}\,,}\label{eom}}
which is valid for spatially smooth magnetic profiles (the so-called adiabatic approximation) and weak currents. Here, $\alpha$ is the Gilbert damping constant, $\beta$ is another dimensionless phenomenological parameter whose physical meaning will be discussed later,  $\mu$ is the electrochemical potential, $\sigma_s$ is the spin-$s$ conductivity (along the local magnetization direction $\mathbf{m}$) relating \textit{particle} currents to $\boldsymbol{\nabla}\mu$, $S$ is the equilibrium spin density of the ferromagnet along $\mathbf{m}$, and $\gamma=M/S$ is the gyromagnetic ratio. Recall that the effective field $\mathbf{H}_{\rm eff}$ is the quantity defined to be \textit{thermodynamically conjugate} to the magnetization: $\mathbf{H}_{\rm eff}=\partial_\mathbf{M}\mathcal{F}$ (note the sign difference from the standard definition), where $\mathcal{F}$ is the free energy and $\partial_\mathbf{M}$ stands for the functional derivative. The other thermodynamic variable we will consider is the electron density $\rho(\mathbf{r},t)$, whose thermodynamically conjugate counterpart is the electrochemical potential $\mu=\partial_\rho\mathcal{F}$.

Suppose we perturb the electron density with respect to an equilibrium with some static magnetic texture and uniform chemical potential, and consider the ensuing magnetic response. Eq.~\rf{eom} then describes the nonequilibrium coupling of the magnetization dynamics to the electron density's thermodynamic conjugate, which is slightly out of equilibrium. The Onsager reciprocity principle\cite{landauBOOKv5} allows us to immediately write down the response of the electron density to a small modulation of the effective field $\mathbf{H}_{\rm eff}$, with respect to an equilibrium configuration. To simplify things, let us for a moment disregard Gilbert damping $\alpha$ in Eq.~\rf{eom} and return to include it later on. An electric response to a magnetic perturbation then becomes \footnote{Note that the Onsager reciprocity relations are somewhat special in the present case: since one of the thermodynamic quantities (namely, the magnetization) is odd under time reversal, the matrix that couples the rates of the relaxation of the involved quantities to their thermodynamic conjugates is antisymmetric, which is subject also to changing the sign of the magnetic field and the equilibrium magnetization.\cite{landauBOOKv5} Hence the overall minus sign in Eq.~\rf{rho} and the sign change in front of $\beta$.}
\eq{\partial_t\rho=-\frac{\gamma\hbar\left(\sigma_\uparrow-\sigma_\downarrow\right)}{2}\nabla_i\left\{\mathbf{H}_{\rm eff}\cdot\left[\left(1+\beta\mathbf{m}\times\right)\nabla_i\mathbf{m}\right]\right\}\,.\label{rho}}
By comparing Eq.~\rf{rho} with the continuity equation $\partial_t\rho=-\nabla_ij_i$, we can identify the particle current as
\eq{j_i=\frac{\gamma\hbar\left(\sigma_\uparrow-\sigma_\downarrow\right)}{2}\mathbf{H}_{\rm eff}\cdot\left[\left(1+\beta\mathbf{m}\times\right)\nabla_i\mathbf{m}\right]\,.\label{j}}
Since for each spin species, $\mathbf{j}_s=\sigma_s\mathbf{F}_s$, where $\mathbf{F}_s$ is the effective force, we finally get for the latter $\mathbf{F}_{\uparrow,\downarrow}=\pm\mathbf{F}$ \footnote{Strictly speaking, the analysis based on Eq.~\rf{eom} can only capture the total particle current \rf{j}. In order to rigorously calculate the spin-resolved forces, we would have to explicitly take into account one more thermodynamic variable, namely, the nonequilibrium spin accumulation (which is proportional to the chemical potential mismatch between the up and down electrons along the local magnetization) and consider its action on the magnetic dynamics. The latter program would push us too much off track, and we choose not to pursue it here. We only wish to note that the $\beta$ term describing the spin-dephasing correction to the fictitious field \rf{Ei} may in general become different for the two spin species when the ferromagnetic exchange energy is comparable to the Fermi energy.}, where
\eq{\al{F_i&=\frac{\hbar}{2}\left(\mathbf{m}\times\partial_t\mathbf{m}\right)\cdot\left[\left(1+\beta\mathbf{m}\times\right)\nabla_i\mathbf{m}\right]\\&=\frac{\hbar}{2}\left[\mathbf{m}\cdot\left(\partial_t\mathbf{m}\times\nabla_i\mathbf{m}\right)+\beta\left(\partial_t\mathbf{m}\cdot\nabla_i\mathbf{m}\right)\right]\,,}\label{Fb}}
after inverting the magnetic equation of motion \rf{eom} in order to express the effective field $\mathbf{H}_{\rm eff}$ in terms of the magnetization dynamics $\mathbf{m}(\mathbf{r},t)$. (Note that since the currents themselves are now generated by the magnetization dynamics, we can neglect their backaction on the magnetic response, when inverting the equation of motion to express $\mathbf{H}_{\rm eff}$ in terms of $\mathbf{m}$, since it would give rise to higher-order terms.) Equation~\rf{Fb} is a key result of this paper. It is also easy to show that taking into account Gilbert damping $\alpha$ has no consequences for the final result \rf{Fb} [after rewriting Eq.~\rf{eom} in the Landau-Lifshitz form, in order to eliminate the $\partial_t$ term on the right-hand side and thus make the equation suitable for the Onsager theorem]. This is not surprising, since the physics of the Gilbert damping $\alpha$ does not have to be related to the magnetization|particle-density coupling that determines the force \rf{Fb}.\cite{tserkovJMMM08}

Physically, the $\beta$ correction in Eq.~\rf{Fb} is related to a slight spin misalignment of electrons propagating through an inhomogeneous magnetic texture with the local direction of the magnetization $\mathbf{m}$. In the limit of $\Delta_{\rm xc}\to\infty$, this misalignment vanishes and so should $\beta$, reducing the result \rf{Fb} to Eq.~\rf{Ei}. Indeed, a microscopic derivation of Eq.~\rf{eom} shows $\beta\sim\hbar/(T_2\Delta_{\rm xc})$, where $T_2$ is the characteristic transverse spin relaxation time.\cite{tserkovJMMM08} The $\beta$ term in Eq.~\rf{Fb} can thus be viewed as a correction to the topological structure of the electron transport rigidly projected on the magnetic texture, due to the remaining transverse spin dynamics and dephasing. Such a $\beta$ correction was first reported in Ref.~\onlinecite{duinePRB08}, which used a very different and more technical language and did not benefit from the reciprocity relation with the current-driven magnetic dynamics \rf{eom}. Our phenomenological derivation of Eq.~\rf{Fb} based on the Onsager theorem provides a much simpler framework for studying these subtle spin-dephasing effects.

\begin{figure}
\centerline{\includegraphics[width=\linewidth]{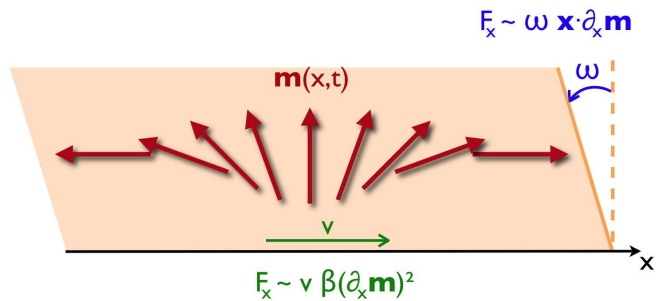}}
\caption{Two simple scenarios for voltage generation by the magnetic dynamics: (1) Magnetic texture $\mathbf{m}(x,t)$, such as a domain wall along the $x$ axis, is steadily rotating around the $x$ axis and (2) the same texture rigidly sliding along the $x$ axis. In the former case, the force $F_x$ acting on electrons is proportional to the frequency of rotation $\omega$, with the dominant term having a purely geometric meaning in terms of the position-dependent Berry-phase accumulation rate. (An alternative physical explanation can also be provided by transforming to the rotating frame of reference and applying the Larmor theorem.) In the case of the sliding dynamics, the leading contribution to the magnetically-induced force is proportional to the spin-dephasing rate (parametrized by $\beta$) and the ``curvature" of the texture profile $(\partial_x\mathbf{m})^2$.}
\label{fig}
\end{figure}

Let us now discuss the measurable consequences of Eq.~\rf{Fb} in two simple scenarios sketched in Fig.~\ref{fig}. Consider a nontrivial one-dimensional magnetic profile along the $x$ axis, such as a magnetic domain wall in a narrow wire, with negligible transverse spin inhomogeneities. First, let us look at a steady rotation of the entire one-dimensional texture around the $x$ axis, with a constant frequency $\omega$. Then, $\partial_t\mathbf{m}=\omega\mathbf{x}\times\mathbf{m}$ and
\eq{\Delta V=-\int dxF_x=-\frac{\hbar\omega}{2}\mathbf{x}\cdot\int\left(d\mathbf{m}+\beta\mathbf{m}\times d\mathbf{m}\right)\,.\label{V1}}
In the absence of spin dephasing $\beta$, this result can be easily understood by transforming into the rotating frame of reference: By the Larmor theorem, this corresponds to a fictitious field along the $x$ axis: $H^\prime=-(\hbar\omega/2)\hat{\sigma}_x$. For spins up (down) projected on the local magnetization direction, this corresponds to the potential $V=\mp(\hbar\omega/2)\mathbf{x}\cdot\mathbf{m}$. It is equally straightforward to interpret this result in terms of the rate of the Berry phase accumulation by spins adiabatically following the steady exchange field precession,\cite{berryPRSLA84,barnesPRL07} which is proportional to the position-dependent solid angle enclosed by spin precession. The $\beta$ term in Eq.~\rf{V1} gives a correction to these idealistic considerations, which depends on the geometry of the magnetic texture. Next, we consider the voltage induced by a rigid translation of a one-dimensional magnetic texture $\mathbf{m}(x-vt)$ along the $x$ direction with velocity $v$. The corresponding force
\eq{F_x=-\frac{\hbar}{2}\beta v\left(\partial_x\mathbf{m}\right)^2}
is then entirely determined by the $\beta$ term, which drags spins down along the direction of the magnetic texture motion and spins up in the opposite direction. This is analogous to the current-driven domain wall velocity in one dimension, which, for smooth walls and low currents, is proportional to $\beta$.\cite{tserkovJMMM08}

Finally, we need to include spin-flip relaxation time $T_1$ and derive spin-charge diffusion equations, enforcing local charge neutrality. Assuming diffusive transport, the force \rf{Fb} can now be added as a contribution to the gradient of the effective electrochemical potential. The diffusion equation for spin-$s$ particles is then given by
\eq{(\partial_t-D_s\nabla^2)\rho_s+\sigma_s\left(s\boldsymbol{\nabla}\cdot\mathbf{F}-\nabla^2V_c\right)=\frac{\rho_{-s}}{\tau_{-s}}-\frac{\rho_{s}}{\tau_{s}}\,,}
where $\rho_s$ is the nonequilibrium (spin-$s$) particle density, $D_s$ is the diffusion coefficient, and $\tau_s$ is the spin-flip time. Recall that the conductivity is related to the density of states $N_s$ by the Einstein's relation: $\sigma_s=N_sD_s$. $V_c$ is the electric potential, which has to be found self-consistently by enforcing local charge neutrality. Note that the equilibrium considerations require that $\tau_s/\tau_{-s}=N_s/N_{-s}$. We should also stress that the force \rf{Fb} may have a finite curl, so that we cannot generally describe it by a fictitious potential. After straightforward manipulations, we can decouple the diffusion equation for the spin accumulation $\mu_\sigma$ (defined as the difference between the spin-up and spin-down electrochemical potentials, divided by 2) from the average electrochemical potential $\mu$ as follows:
\eq{\al{\left(\partial_t+\tau^{-1}-D\nabla^2\right)\mu_\sigma&=-D\boldsymbol{\nabla}\cdot\mathbf{F}\,,\\-\nabla^2\mu&=P\left(\nabla^2\mu_\sigma-\boldsymbol{\nabla}\cdot\mathbf{F}\right)\,.}\label{SD}}
Here, $P=(\sigma_\uparrow-\sigma_\downarrow)/(\sigma_\uparrow+\sigma_\downarrow)$ is the conductivity polarization, $D=(D_\uparrow+D_\downarrow)/2-P(D_\uparrow-D_\downarrow)/2$ is the effective spin-diffusion constant, and $\tau^{-1}=\tau_\uparrow^{-1}+\tau_\downarrow^{-1}$ is the characteristic $T_1^{-1}$ rate for spin flipping. If $\boldsymbol{\nabla}\times\mathbf{F}=0$, we can integrate the second equation to express the electrochemical potential gradient in terms of the force $\mathbf{F}$ and the spin accumulation gradient as follows:
\eq{\boldsymbol{\nabla}\mu=P(\mathbf{F}-\boldsymbol{\nabla}\mu_\sigma)\,,\label{mu}}
assuming the appropriate boundary conditions. According to Eqs.~\rf{SD}, the spin accumulation decays in the absence of the force $\mathbf{F}$ on the scale of the spin-diffusion length $\lambda_{\rm sd}=\sqrt{D\tau}$. Away from the dynamic magnetic texture (on the scale of $\lambda_{\rm sd}$), the generated electrochemical potential \rf{mu} will then be determined simply by integrating the force $\mathbf{F}$. In general, however, especially when $\boldsymbol{\nabla}\times\mathbf{F}\neq0$, one has to revert to Eqs.~\rf{SD}.

In summary, we theoretically studied electron transport generated by a dynamic magnetization texture. We reproduced and generalized the results that recently appeared in literature,\cite{barnesPRL07,duinePRB08,yangCM07} revealing an intricate connection with the theory of the current-induced magnetization dynamics.\cite{tserkovJMMM08} We expect that in practice it is considerably simpler to solve this reciprocal problem, especially for including subtle corrections to the topological Berry phase structure of spins assumed to rigidly follow the time-dependent magnetic profile.

\end{document}